\documentstyle[preprint,aps]{revtex}
\input epsf.tex
\def\DESepsf(#1 width #2){\epsfxsize=#2 \epsfbox{#1}}
\begin{document}
\preprint{\vbox{\hbox{OITS-582}}}
\draft
\title{Importance of Dipole Penguin Operator in $B$ Decays
\footnote{Work supported in part by the Department of Energy Grant No.
DE-FG06-85ER40224.}}
\author{$^1$N.G. Deshpande,
$^1$Xiao-Gang He and $^2$J. Trampeti\' c}
\address{$^1${Institute of Theoretical Science\\
University of Oregon\\
Eugene, OR 97403-5203, USA} \\ $^2${Department of Theoretical
Physics, R. Bo\v{s}kovi\'{c}
 Institute, \\
 P.O.Box 1016, 41001 Zagreb, Croatia}}
\date{September, 1995}
\maketitle
\begin{abstract}
The importance of dipole penguin operator,
$O_{11} = (g_s/8\pi^2)m_b \bar s \sigma_{\mu\nu}RT^a b G^{\mu\nu}_a$,
on $B$ decays is demonstrated. It is shown that branching ratios for
the inclusive decay $b\rightarrow s s\bar s$,
the semi-inclusive decay $b\rightarrow s\phi$, and the exclusive decay
$B\rightarrow \pi K$ are enhanced  by 20\% to 30\%. A useful consequence of
this calculation is the inclusive $b\rightarrow ss\bar s$ spectrum which will
be essential in isolating pure penguin process.

\end{abstract}
\newpage

Flavor changing one loop processes mediated by a gluon play an important role
in understanding the Standard Model (SM)\cite{1,2,3,4,5,6,7}. It is
by now well known that substantial corrections arise in these estimates from
inclusion of the full electroweak effects\cite{2,4,5,6,7} (the $\gamma$, $Z$
penguins and the ``box'' contributions).
Naively, one would think that in the $B$ system electroweak corrections to
gluon exchange may be negligible. This is, in fact, incorrect. The large mass
of the top quark increases the relative contributions of the electroweak
corrections. Our recent estimates for $b\rightarrow s\phi$ process suggests
a reduction in the rate of 20\%$\sim$ 30\%\cite{4,5}.
In this paper we study another contribution, which has been mostly
neglected so far, to nonleptonic B decays. This is the contribution from the
Dipole Penguin Operators (DPO). Similar operators have been shown to
give non-negligible contribution to $\epsilon^{\prime}/\epsilon$\cite{8,9}
and hyperon
decays\cite{9}. We consider three different types of decays in increasing
model dependence, the inclusive
$b\rightarrow sq\bar q$ and $b\rightarrow s s \bar s$, the semi-inclusive
$b\rightarrow s\phi$, and
the exclusive $B^-\rightarrow \bar K^0 \pi^-$ decyas.
We find that the DPO gives
important additional contributions
to these decays and enhances their branching ratios by up to 30\%.
Any future quantitative analysis of B decays that relies on full electroweak
Hamiltonian
will have to include the DPO contributions. An important consequence of this
calculation is the inclusive $b\rightarrow ss\bar s$ spectrum which will be
useful in isolating pure penguin process.

\noindent
{\bf The effective Hamiltonian }

The effective Hamiltonian up to one loop level in electroweak interaction
for charmless $B$ decays can be written as
\begin{eqnarray}
H_{\Delta B=1} = {4G_F\over \sqrt{2}}[V_{ub}V^*_{uq}(c_1O_1 + c_2 O_2)
 - V_{tb}V^*_{tq}\sum^{12}_{i=3} c_iO_i] +H.C.\;,
\end{eqnarray}
where $O_i$ are defined as
\begin{eqnarray}
O_1 = \bar q_\alpha \gamma_\mu L f_\beta\bar
f_\beta\gamma^\mu L b_\alpha\;&,&\;\;\;\;
O_2 = \bar q \gamma_\mu L f\bar
f\gamma^\mu L b\;,\nonumber\\
O_{3(5)} = \bar q \gamma_\mu L b \sum_{q'}
\bar q' \gamma_\mu L(R) q'\;&,&\;\;\;\;
Q_{4(6)} = \bar q_\alpha \gamma_\mu L b_\beta \sum_{q'}
\bar q'_\beta \gamma_\mu L(R) q'_\alpha\;,\nonumber\\
O_{7(9)} ={3\over 2}\bar q \gamma_\mu L b \sum_{q'} e_{q'}\bar q'
\gamma^\mu R(L)q'\;&,&\;\;
Q_{8(10)} = {3\over 2}\bar q_\alpha \gamma_\mu L b_\beta \sum_{q'}
e_{q'}\bar q'_\beta \gamma_\mu R(L) q'_\alpha\;,\nonumber\\
O_{11} ={g_{s}\over{32\pi^2}}m_{b}\bar q \sigma_{\mu \nu} RT_{a}b
G_{a}^{\mu \nu} \;&,&\;\;
Q_{12} = {e\over{32\pi^2}} m_{b}\bar q \sigma_{\mu \nu}R b
F^{\mu \nu} \;,\nonumber
\end{eqnarray}
where $L(R) = (1\mp \gamma_5)/2$, $f$ can be $u$ or $c$, $q$ can be
$d$ or $s$, $q'$ is
summed over $u$, $d$, $s$, and $c$. $\alpha$ and $\beta$ are the color
indices.
$T^a$ is the SU(3) generator with the normalization
$Tr(T^aT^b) = \delta^{ab}/2$. $G^{\mu\nu}_a$ and
$F_{\mu\nu}$ are
the gluon and photon field strength, respectively. $O_2$, $O_1$
are the tree level
and QCD corrected operators. $O_{3-6}$ are the strong
gluon induced operators.  $O_{7-10}$
are the electroweak penguin operators due to $\gamma$ and $Z$
exchange, and ``box'' diagrams at loop level. $O_{11,12}$ are the DPOs.

The Wilson Coefficients (WC) $c_i$ at a particular scale $\mu $
are obtained by first
calculating the WCs at $ m_W$ scale and then
using
the renormalization group equation to evolve them to $\mu$. We have carryed
out this analysis using the next-to-leading order QCD corrected WCs following
Ref\cite{10}. Using $\alpha_s(m_Z)=0.118$, $\alpha_{em}(m_Z)
= 1/128$, $m_t = 176$ GeV and $\mu \approx m_b = 5$ GeV, we obtain\cite{4}
\begin{eqnarray}
c_1 &=& -0.3125\;, \;\;c_2 = 1.1502\;, \;\;c_3 = 0.0174\;, \;\;c_4 = -0.0373\;,
\nonumber\\
c_5 &=& 0.0104\;,
\;\;c_6 = -0.0459\;,\;\; c_7 = 1.398\times 10^{-5}\;,\nonumber\\
 c_8 &=& 3.919\times 10^{-4}\;,
\;\;c_9 = -0.0103\;, \;\;c_{10} = 1.987\times 10^{-3}\;.
\end{eqnarray}
The DPO coefficients at the two loop level
have the following values\cite{11}:
\begin{eqnarray}
c_{11} &=& -0.299\;, \;\;c_{12} = - 0.634\;.
\end{eqnarray}

The above Hamiltonian allows one to study the full set of charmless
nonleptonic  B meson decays.
It is clear that only in $b\rightarrow s$ transitions, the DPO may
be important because of the enhancement factor
$|V_{tb}V_{ts}^*/V_{ub}V_{us}^*|\approx 57$
in Eq.(1)
compared with the tree contribution. The three processes mentioned before
are all of this type. There are two types of DPO contributions. One
from $O_{11}$ and the other from $O_{12}$.
The contribution from $O_{12}$ is suppressed by a
factor of $\alpha_{em}/\alpha_s$ compared with the contributions from $O_{11}$.
We will therefore neglect its contribution in the following analysis.

\noindent{\bf The inclusive decay: $b\rightarrow s q\bar q$}

Let us first consider $b\rightarrow s s \bar s$. This is
a pure penguin induced process and does not involve uncertainties
accosiated
with hadronic matrix elements. It serves as a clear indication of the
importance of DPO. It also might be feasible to isolate this process
experimentally by selecting energetic kaons.
In this process the gluon in the
DPO has to break up into a pair of $s\bar s$.
In our calculation we will use the four-quark operator generated by
$O_{11}$,
\begin{eqnarray}
H_{11} = i{G_F\over \sqrt{2}}{\alpha_s\over \pi k^2}
V_{tb}V_{ts}^* c_{11}m_b \bar s \sigma_{\mu\nu}RT^ab \bar q \gamma^\mu
T^a q k^\nu\;,
\end{eqnarray}
where $k = p_b -p_s$ in the gluon momentum, and $q = s$ for
$b\rightarrow s s\bar s$.

Since there are two s
quarks in $b\rightarrow s s \bar s$ decay, we have to anti-symmetrize the
amplitude. We have
\begin{eqnarray}
A(b(p_b)&\rightarrow& s_1(p_1) s_2(p_2)\bar s(p_s)) = -{G_F\over \sqrt{2}}
V_{tb}V_{ts}^*[ A_{12} \bar s_1\gamma_\mu L b \bar s_2\gamma^\mu L s\nonumber\\
&+&B_{12}\bar s_1\gamma_\mu L b \bar s_2\gamma^\mu R s
+ C_{12}\bar s_1^\alpha L b_\beta \bar s_2^\beta R s_\beta \nonumber\\
&+&D_{12}(p_b+p_1)_\mu\bar s_1 RT^ab \bar s_2 \gamma^\mu T^a s -
(1 \leftrightarrow 2) ]\;,
\end{eqnarray}
where
\begin{eqnarray}
A_{12}& =& 4c_3 +4c_4 - 2c_9 -2c_{10} -{\alpha_sc_{11}m_b^2\over 2 \pi}
({1\over N_c}{1\over s}-{1\over t})\;,\nonumber\\
B_{12} &=& 4c_5-c_7-{\alpha_sc_{11}m_b^2\over 2\pi}{1\over N_c s}\;,\nonumber\\
C_{12}&=&-2(4c_6-2c_8 +{\alpha_sc_{11}m_b^2\over 2\pi}{1\over t})\;,\nonumber\\
D_{12} &=& -{\alpha_sc_{11}m_b\over \pi}{1\over s}\;,
\end{eqnarray}
where $s = (p_b-p_1)^2$, $t = (p_b-p_2)^2$, and $N_c = 3$ is the number of
colors.
The coefficients $(A_{21},B_{21},C_{21},D_{21})$ are obtained by exchanging $s$
and $t$ in the
above equations. The decay width is given by
\begin{eqnarray}
\Gamma = {1\over (2\pi)^3}{1\over 32 m_b^3}
\int ds dt |A|^2\;,
\end{eqnarray}
where the integration limits for $s$ are integrated $4m_s^2$ and $(m_b-m_s)^2$,
and the limits for $t$ are  $(m_b^2-s+3m_s^2 \pm \sqrt{(1-4m_s^2/s)
((m_b^2-s-m_s^2)^2-4sm_s^2)})/2$. The integrand $|A|^2$ is
given by
\begin{eqnarray}
|A|^2 &=&{1\over 2N_c2!}{G_F^2\over 2}|V_{tb}V_{ts}^*|^2\left \{
[4N_c^2(|A_{12}|^2+|A_{21}|^2)
+8N_cRe(A_{12}A_{21}^*)](s+t)(m_b^2-s-t)
\right .\nonumber\\
&+&N_c^2[(4|B_{12}|^2+|C_{21}|^2)t(m_b^2-t) +
(4|B_{21}|^2+|C_{12}|^2)s(m_b^2-s)]\nonumber\\
&+&2(N_c^2-1)[|D_{12}|^2t(m_b^2-s)
+|D_{21}|^2s(m_b^2-t)](m_b^2-s-t)\nonumber\\
&-&2{N_c^2-1\over N_c}Re(D_{12}D_{21}^*) s t(m_b^2-s-t)\nonumber\\
&-&4N_c[Re(B_{12}C_{21}^*) t (m_b^2-t) +Re(C_{12}B_{21}^*) s
(m_b^2-s)]\nonumber\\
&+&2(N_c^2-1)m_b[Re(2A_{12}D_{21}^* - C_{12}D_{21}^*)s\nonumber\\
&+&Re(2D_{12}A_{21}^*-D_{12}C_{21}^*)t] (m_b^2-s-t)\left .\right \}\;.
\end{eqnarray}
Here we have averaged the initial spin and color factors, and also
taken into account the symetrization factor $2!$.

In our numerical evaluation, we will use $|V_{ts}| \approx |V_{cb}| = 0.04$,
$\alpha_s(m_b) = 0.25$, $m_b = 5$GeV.
Since s quarks hadronize into kaons, it seems appropriate to use the
constituent mass. We use $m_s = 0.5$ GeV. Using these numbers, we calculate the
relative branching ratio $R = \Gamma(b\rightarrow s s\bar
s)/\Gamma(b\rightarrow
c e\bar \nu_e)$.  Without the DPO effects, $R = 0.018$. When DPO effects are
included,
$R = 0.024$. We see that the DPO enhances the branching ratio by about 30\% and
should not be ignored. This result improves upon the result
without QCD corrections discussed in Ref.\cite{12}.

It is essential to know the distribution of events with respect to invariant
masses for selecting the desired events against background.  To this end
we present two useful spectra. In Figures 1 and 2, we plot the normalized
event distributions with
respect to diquark invariant mass
$x = (p_1+p_2)^2/(m_b-m_s)^2$ and quark anti-quark invariant mass
$y = (p_1+p_s)^2/(m_b-m_s)^2$. Although the effect of DPO on the overall
branching ratio is significant,
we see that its effect on the normalized distributions is very
small.

We now consider  $b\rightarrow s q\bar q$ with $q\bar q$ summed over
$u \bar u$, $d\bar d$ and $s\bar s$. This process may be easier
to measure experimentally because this process can be isolated by selecting
single kaons having high meneta.  In this calculation we do not need to
antisymmetrize the amplitudes for $b\rightarrow s u\bar u$ and
$b\rightarrow s d\bar d$. We again use constituent masses with
$m_u = m_d = 0.3$ GeV and $m_s = 0.5$ GeV. We find that
the effect of DPO becomes weaker. It enhances the branching ratio by
about 10\% compared with the branching ratio without DPO contribution.
The event distribution (shown in Fig 2) with respect to the $q\bar q$
invariant mass $y = (p_q+p_{\bar q})^2/(m_b-m_s)^2$ is similar to
$b\rightarrow ss\bar s$. The DPO effect on the ditribution is negligible.
The ratio
$\bar R = \Gamma(b\rightarrow sq\bar q)/\Gamma(b\rightarrow c e\bar\nu_e)$ now
also depends on the
$b\rightarrow u$ tree operators ($O_{1,2}$) which involves the weak phase
$\gamma$.
We find $\bar R$ is between $0.11\sim 0.6$ when varying the angle $\gamma$
between $0^0 \sim 180^0$.

\noindent
{\bf The semi-inclusive decay: $b\rightarrow s\phi$}

{}From experience we know that processes involving hadronic matrix element are
difficult to handle because there is no satisfactory method to calculate these
matrix elements at present. To have an idea how important the DPO effects are,
we use factorization approximation to evaluate the DPO contribution to
$b\rightarrow s\phi$.

In the factorization approximation, the effective Hamiltonian in Eq.(4) gives
vanishing matrix element without Fierz transformation because $s\bar s$
operator is in color octect states. Only the Fierz transformed Hamiltonian
contributes.  After a Fierz transformation, we obtain
\begin{eqnarray}
H_{eff, F}& &= -{G_F\over \sqrt{2}}{\alpha_s\over 4\pi k^2}
V_{tb}V_{ts}^* {N_c^2-1\over N_c^2}c_{11}m_b
[2m_b \bar s^\alpha\gamma_\mu L s_\beta\bar s^{\alpha'}\gamma^\mu L b_{\beta'}
\nonumber\\
&-&4 m_b \bar s^\alpha R s_\beta\bar s^{\alpha'} L b_{\beta'}
+(p_b^\mu+p_s^\mu)
(\bar s^\alpha \gamma_\mu L s_\beta \bar s^{\alpha'} R b_{\beta'}
+ \bar s^\alpha R s_\beta \bar s^{\alpha'} \gamma_\mu R b_{\beta'}\nonumber\\
&-&i \bar s^\alpha \sigma_{\mu \nu}R s_\beta \bar s^{\alpha'}\gamma^\nu R
b_{\beta'}
+i\bar s^\alpha \gamma^\nu L s_\beta \bar s^{\alpha'} \sigma_{\mu\nu} R
b_{\beta'})]
(\delta_{\alpha}^\beta \delta_{\alpha'}^{\beta'}
- {2N_c\over N_c^2-1} T^{a\beta}_{\;\alpha} T^{a\beta'}_{\;\alpha'})\;.
\end{eqnarray}
The second term in the last bracket gives vanishing contribution in the
factorization approximation. We have
\begin{eqnarray}
A(b\rightarrow s \phi) &=&
-{G_F\over \sqrt{2}}{\alpha_s\over 16\pi k^2}
V_{tb}V_{ts}^*{N_c^2-1\over N_c^2} c_{11}m_big_\phi \epsilon_\mu [ m_b\bar s
\gamma^\mu L b +
4(1+{m_s^2\over m_\phi^2}) p_b^\mu \bar s R b\nonumber\\
&-& m_s(1-2{m_b^2+2m_\phi^2-m_s\over m_\phi^2})\bar s \gamma^\mu R b
-4{m_sm_b\over m_\phi^2}p_b^\mu\bar s L b)\;,
\end{eqnarray}
where $\epsilon_\mu$ is the polarization vector of $\phi$.
In the above, we have used the parametrization $<\phi|\bar s \gamma_\mu s|0>=
ig_\phi \epsilon_\mu$, with $g_\phi^2 = 0.0586$ GeV$^4$;
and $<\phi|\bar s \sigma^{\mu\nu}s|0> = \delta
(\epsilon^\mu p_\phi^\nu-p_\phi^\mu \epsilon^\nu)$ with the quark model
relation $\delta = -2m_sg_\phi/m_\phi^2$.  We assumed that the $s$ and
$\bar s$ quarks inside $\phi$ each
have similar momentum $p_\phi/2$. In this approximation,
$k^2 = (m_b^2-m_\phi^2/2+m_s^2)/2$. Using the approximation $A(b\rightarrow
s\phi)
\approx A(B\rightarrow X_s \phi)$,
for $N_c = 3$ we find that without the DPO contribution the branching ratio
$BR(B\rightarrow
X_s \phi)$ is  $1.0\times 10^{-4}$. With the DPO contribution,
$BR(B\rightarrow X_s \phi) = 1.2\times 10^{-4}$. The branching ratio is
enhanced
by about 20\% when the DPO contribution is included.  We also checked
the DPO effect with $N_c = 2$ favored by phenomenological
fitting of the experimental data\cite{nc}. In this case the branching ratio is
about
two times larger, but the DPO contribution only enhances the braching ratio
by about 10\%.

\noindent
{\bf The exclusive decay: $B\rightarrow K\pi$}

Here we carry out a calculation for  $B^-\rightarrow \pi^-\bar K^0$ as an
example.
We, again, use the factorization approximation to estimate the amplitude. We
have
\begin{eqnarray}
A_{11}(B^-&\rightarrow&\pi^-\bar K^0) =
-{G_F\over \sqrt{2}}{\alpha_s\over 4\pi k^2}V_{tb}V_{ts}^*{{N^2_c -1}
\over{N^2_c}}c_{11}m_b\nonumber\\
&\times&[2m_{b}
<\bar K^0|\bar s
\gamma_{\mu}L d|0> <\pi^-|\bar q\gamma_{\mu} L b|B^->\nonumber\\
&-&4m_{b}<\bar K^0|\bar s Rd|0>
<\pi^-|\bar d L b|B^-> \nonumber \\
& + & (p_b + p_s)_{\mu}(<\bar K^0|\bar s
\gamma_{\mu} L d|0><\pi^-|\bar d  R b|B^->\nonumber\\
&+& <\bar K^0|\bar s R d|0>
<\pi^-|\bar d \gamma_{\mu} R b|B^-> \nonumber \\
& - &i<\bar K^0|\bar s \sigma_{\nu \mu} R d|0><\pi^-| \bar d
\gamma_{\nu} R b|B^-> \nonumber\\
&+& i<\bar K^0|\bar s \gamma_{\nu} L d|0><\pi^-|
\bar d \sigma_{\nu \mu} R b|B^->)\;.
\end{eqnarray}
In the above
we have neglected the annihilation contributions of the form $<\pi^-\bar
K^0|\bar s
\Gamma_1 u|0><0|\bar u\Gamma_2|b|B^->$, where
$\Gamma_{i}$ indicate the appropriate gamma matrices
with corresponding Lorentz indices. These contributions are generally
much smaller than the ones considered here\cite{6}.

We proceed with the
Lorentz decomposition of the typical hadronic matrix elements:
\begin{eqnarray}
&<&\bar K^{0}\left|\overline{s}\gamma^{\mu}\gamma_{5}d\right|0> =
iq^{\mu}f_{K}\;,\;\;\mbox{with}\; f_K = 160\;\mbox{MeV}\;,\nonumber\\
&<&\pi^{-}\left|\overline{d}\gamma^{\mu}(1\pm \gamma_{5})b\right|B^{-}>=
(p_B+p_\pi)^{\mu}f^+ + (p_B^{\mu}-p_\pi^\mu)f^-\;,\nonumber\\
&<&\pi^{-}\left|\overline{d}\sigma^{\mu \nu}(1+\gamma_5)b\right|B^{-}>=
-i2h(p_B^\mu p_\pi^\nu - p_B^\nu p_\pi^\mu )\;.
\end{eqnarray}
We use the Heavy Quark Effective Theory (HQET) relation\cite{13} $h =
(f^+ - f^-)/4m_b$ to relate $h$ to $f^{\pm}$.

In terms of the decay constant $f_K$ and form factors $f^\pm$, $h$, the
$B^-\rightarrow \bar K^0 \pi^-$
decay amplitude is given by
\begin{eqnarray}
A_{11}(B^-&\rightarrow& \bar K^0 \pi^-) = i{G_F\over \sqrt{2}}
{\alpha_s\over{4\pi k^2}}V_{tb}V^*_{ts}c_{11}m_b
{{1-N^2_c}\over{N^2_c}}f_K \left[\left((m^{2}_{B}-m^{2}_{\pi})f_{B\pi}^+ +
m^{2}_{K} f_{B\pi}^-\right) \right.\nonumber \\
& \times &
\left({m_{b}\over 2} + {{m_b m^2_K}\over{(m_b-m_d)(m_s+m_d)}} +
{{m^2_B +2m^2_K -m^2_\pi}\over{8(m_b-m_d)}}\right) \nonumber \\
& - &
{{m^2_K}\over{4(m_s+m_d)}}\left((2m^{2}_{B}-{1\over2}m^{2}_{K})
f_{B\pi}^+ +
{1\over2}\left( m^{2}_{B}+2m^{2}_{K}-m^2_\pi\right) f_{B\pi}^-\right)
\nonumber\\
& + &
{1\over8}h_{B\pi}\left(m^4_B+m^4_K+m^4_\pi -
2m^{2}_{B}m^{2}_{K}-2m^{2}_{B}m^2_\pi-2m^{2}_{K}m^2_\pi\right)\left.\right]\;.
\end{eqnarray}

In the above we have used  the same approximation for the kinemetics as
 in the case of $b\rightarrow s\phi$. We assume that the two quarks s and
d inside $\bar K^0$ share the meson momentum
equally, and the b quark carries all the $B$ meson
momentum. In this case $k^2 = (m^2_B-m^2_K/2+m^2_\pi)/2$

Using the form factors evaluated in Ref.\cite{14}, we obtain the branching
ratio
$BR(B^-\rightarrow \pi^-\bar K^0)$ to be $0.83\times 10^{-5}$ without the DPO
contribution. When
the DPO contribution is included, $BR(B^-\rightarrow \pi^-\bar K^0) =
1.0\times 10^{-5}$.
Again, the DPO enhances the branching ratio by about 20\%. The dependence on
$N_c$ is weak in this case.

{}From the above discussions we clearly see that the DPO has important
effects on the branching ratio of B decays. These effects enhances the
branching
ratios by 20\% to 30\%. All analyses which involve numerical values of the
strong penguin contribution
are affected. These analyses need to be re-examined.
However, since the DPO has the same transformation properties under
$SU(3)$ flavor (and Isospin) symmetry as the ordinary strong penguin operators,
many of the analyses related to CP violation in $B$ decays
which only depends on transformation properties under
flavour symmetry\cite{6,15,16} and processes where the strong penguin does not
contribute \cite{5},
will not be affected.

\begin{figure}[htb]
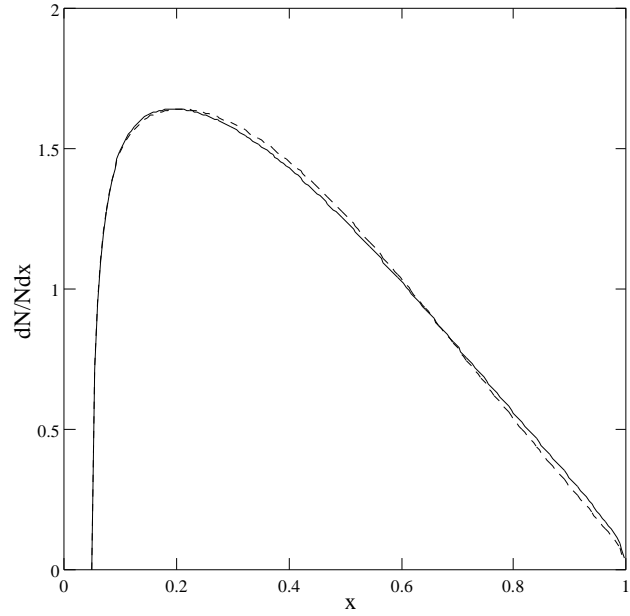

\centerline{ \DESepsf(ss.epsf width 8 cm) }
\smallskip
\caption {The solid and dashed lines are for the event distribution
with respect to the normalized diquark invariant mass x with and without the
dipole
penguin operator contribution, respectively.}
\label{gamma}
\end{figure}

\begin{figure}[htb]
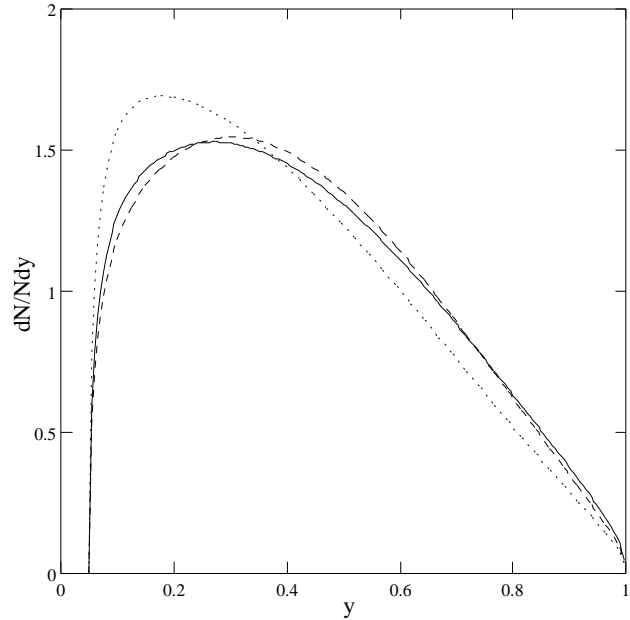

\centerline{ \DESepsf(ssbar.epsf width 8 cm) }
\smallskip
\caption {The solid and dashed lines are for the event distribution
with respect to normalized quark anti-quark invariant
mass y for $b\rightarrow ss\bar s$
with and without the dipole
penguin operator contribution, respectively. The dotted line is for
$b\rightarrow s q\bar q$ with dipole penguin operator contribution.}
\label{gamma}
\end{figure}

\end{document}